\documentclass[12pt]{article} \textheight=24 true cm
\usepackage{graphicx}
\usepackage{amsmath}
\usepackage{amssymb}

\usepackage[colorlinks]{hyperref}
\begin{document}
\begin{center}
{\Large\bf General relativistic plasma in higher dimensional space
time
}\\[8 mm]

D. Panigrahi \footnote {Relativity and Cosmology Research Centre,
Jadavpur University, Kolkata - 700032, India ,
 e-mail: dibyendupanigrahi@yahoo.co.in ,
Permanent Address : Kandi Raj College, Kandi, Murshidabad 742137,
India}
  and S. Chatterjee\footnote{Relativity and Cosmology Research Centre, Jadavpur University,
Kolkata - 700032, India, and also at NSOU, New Alipore College,
Kolkata  700053,
e-mail : chat\_ sujit1@yahoo.com\\Correspondence to : S. Chatterjee} \\[6mm]

\end{center}
\begin{abstract}
The well known (3+1) decomposition of Thorne and Macdonald is
invoked to write down the Einstein-Maxwell equations generalised
to (d+1) dimensions and also to formulate the plasma equations in
a flat FRW like spacetime in higher dimensions (HD). Assuming an
equation of state for the background metric we find solutions as
also dispersion relations in different regimes of the universe in
a unified manner both for magnetised(un) cold plasma. We find that
for a free photon in expanding background we get maximum redshift
in 4D spacetime, while for a particular dimension it is so in pre
recombination era. Further wave propagation in magnetised plasma
is possible for a restricted frequency range only, depending on
the number of dimensions. Relevant to point out that unlike the
special relativistic result this allowed range evolves with time.
Interestingly the dielectric constant of the plasma media remains
constant, not sharing the expansion of the background, which
generalises a similar 4D result of Holcomb-Tajima in radiation
background to the case of higher dimensions with cosmic matter
obeying an equation of state . Further, analogous to the flat
space static case we observe the phenomenon of Faraday rotation in
higher dimensional case also.
\end{abstract}
   ~~~~~~~KEYWORDS : cosmology; higher dimensions; plasma

~~~PACS :   04.20, 04.50 +h

\section*{1. Introduction}

While great strides have been made by general relativists to
address the issues coming out of the recent observations in the
field of astrophysics and cosmology and despite the fact that more
than 90 percent of the cosmic stuff in  stellar interior and
intergalactic spaces is made up of matter in plasma state the much
sought after union between the plasma dynamics and general
relativity still remains elusive. Although we occasionally come
across stray works like `plasma suppression of large scale
structure formation in the universe' ~\cite{prl} as also even the
simulation techniques ~\cite{smoot} where it is argued that the
geometry of the spacetime can be purported to be structured by
observing sound waves in primordial plasmas the fact remains that
as the field equations both in general relativity and plasma
dynamics are highly nonlinear it is very difficult to obtain
closed form solutions in physics. So either numerical relativity
or a linearised approximation of the plasma equations is
preferred, before going in for more complicated non linear
phenomena. If we briefly trace the thermal history of the universe
theorists believe that from approximately $t= 10^{-3}$ s to $t=1$
s at temperatures $T > 10^{10}$ K there was an electron positron
plasma at ultra relativistic temperatures. With cooling the plasma
comprises mainly electron and hydrogen ions, may be with small
amount of helium and other light elements in thermal equilibrium
with photons. This period has come to be known as radiation
dominated era because it is believed that the energy density of
photons much exceed that of matter ~\cite{bb}. As the temperature
decreased further with expansion at around $t \sim 10^{13}$ s
recombination of ionized hydrogen atoms took place with consequent
decoupling of matter and radiation and at a certain stage the
universe becomes matter dominated. Consequently the temporal
behaviour of the scale factor correspondingly changes with the
evolution of matter at different time scales. Nevertheless, the
studies of the effects of this expansion on the electromagnetic
interactions of the matter, including the longitudinal and
transverse modes have not, so far got the legitimate attention it
deserves.

Another area of current interest is the role of an external
magnetic field in many astrophysical systems and cosmology and
possible sources for the field in different scales
~\cite{Kronberg} although the origin of the field continues to
evade plausible physical explanations ~\cite{giovanni} so far.
Granted that temperature (particularly at early universe) is a
known enemy of magnetism there are no compelling reasons why
magnetic fields should not have been present in the early Universe
either. Indeed, the presence of large scales magnetic fields in
our observed Universe is a well established experimental fact.
Since their first evidence in diffuse astrophysical plasmas beyond
the solar corona ~\cite{Kronberg,giovanni} magnetic fields have
been detected in our galaxy and in our local group through Zeeman
splitting and through Faraday rotation measurements of linearly
polarized radio waves. The Milky Way possesses a magnetic field
whose strength is of the order of the micro\-gauss corresponding
to an energy density roughly comparable with the energy density
today stored in the Cosmic Microwave Background Radiation (CMB\-R)
energy spectrum peaked around a frequency of 30 G\-Hz. Faraday
rotation measurements of radio waves from extra-galactic sources
also suggest that various spiral galaxies are endowed with
magnetic fields whose intensities are of the same order as that of
the Milky Way. The existence of magnetic fields at even larger
scales (intergalactic scale, present horizon scale, etc.) cannot
be excluded, but it is still quite debatable since, in principle,
dispersion measurements (which estimate the electron density along
the line of sight) cannot be applied in the intergalactic medium
because of the absence of pulsar signals. Given the fact that a
seed field does exist its amplification mechanism through the so
called dynamo effect is relatively well understood ~\cite{beck}
for an expanding cosmological model with the help of
magneto\-hydrodynamical (M\-HD) equations ~\cite{barrow}. On the
other hand when discussing a primordial magnetic field one should
also bear in mind that a large value would create significant
anisotropy of the background geometry ~\cite{zeldovich} with
consequent impact on CMB\-R findings. As the large scale isotropy
of the CMB\-R fairly excludes that possibility there should be
efficient mechanism to rapidly damp that field.

On the other hand there has been, of late, a resurgence of
interests in physics in higher dimensional spacetime ~\cite{pr} in
its attempts to unify all the forces in nature, to give a physical
explanation of the current accelerating era of the universe
without bringing in any hypothetical quintessencial type of scalar
field ~\cite{gr} by hand, in the newly fashionable area of brane
cosmology ~\cite{subenoy} where the gravity is supposed to act in
the bulk while other forces in the physical 3D space. It has also
received serious attention in the recently proposed induced matter
theory pioneered by Wesson and others~\cite{wesson}. Most
importantly both higher dimensional spacetime and cosmological
plasma have one thing in common -\emph{ both  are very relevant in
the context of early universe}. While early universe may be
loosely viewed  as  the history of evolution of matter in plasma
state it can also be shown that starting from a higher dimensional
phase the Einstein's generalized field equations dictate results
such that as time evolves the 3D space expands while the extra
dimensions shrink till plackian length when some stabilizing
mechanism ( for example, quantum gravity, casimir effect, a
repulsive potential) halts the shrinkage down to a very small
length, say planckian size as to be invisible with the low energy
physics at the moment. So the world around us appears manifestly
three dimensional. But it should be emphasized that the time
scales for the spontaneous self compacti\-fication (SSC) and the
onset of nucleosynthesis clearly differ with SSC occurring much
earlier. So many of our findings in MHD lose much of their
relevance when working in higher dimensional spacetime. While
literature abounds with works on the effects of the expanding
background on the matter distribution and vice versa as also on a
large number of other physical processes scant attention has been
paid so far to address the issues resulting from the expanding
universe on the propagation of say, electromagnetic wave as also
its interactions in a plasma media. To authors' knowledge Holcomb
and Tajima (HT)~\cite{Hol,Holc}, Banerjee et al ~\cite{absc} and
later Dettmann et al ~\cite{dettmann} made important contributions
in this regard. HT investigated the electromagnetic wave
propagation in a radiation dominated and later in a matter
dominated background both with or without any plasma material and
also generalised it to the magnetized case both warm and cold. For
the ultra relativistic case it is observed that all the modes
redshift at the same rate i.e., photons are, so to say, self
similar. Though not explicitly pointed out in their works we think
this, however, may be a direct consequence of the conformal
flatness of the FRW metric chosen by them. Later Banerjee et al
discussed this in a little more general way. Dettmann et al got
the similar results starting from a kinetic theory approach. HT's
findings for the matter dominated model, however, differ sharply
from the first paper in the sense that while in the unmagnetised
plasma case the photons redshift identically but here for the
Alfven waves the frequency redshifts in a bizarre fashion unlike
the case of free photons, depending on the magnitudes of the
plasma density and also strength of the external magnetic field.
On the other hand Dettmann shows that even for the unmagnetised
plasma the plasma oscillations and the photons do not share the
identical temporal dependence if they are decoupled. They,
however, treated the whole situation from kinetic energy
considerations. In the present work we have investigated the
plasma dynamics in an expanding higher dimensional background in a
very general way. Taking a (d+1) dimensional flat FRW type of
metric as background, which one of us ~\cite{sc1} derived earlier
in a different context we first take the case of propagation of an
electromagnetic wave in vacuum. To make things very general we
assume an equation of state for the background as $p =\gamma\rho$.
Taking $\gamma = \frac{1}{d}$, (as we are dealing with a (d+1)
dimensional case) for the radiation and $\gamma= 0$, for the dust
case in the resulting solution we observe that solutions are very
similar to the special relativistic form except that the
frequencies red shift depending on $\gamma$ and the field
amplitude is no longer a constant decaying with the expansion
rate, reminiscent of the acoustic case where the damping occurs
due to some form of dissipating mechanism. Here the expansion of
the background, in a sense, takes the role of dissipation, causing
this type of damping. It is observed that with number of
dimensions the red shift decreases, being maximum in the usual 4D
case. Moreover, red shift is less in the post recombination era
compared to the early universe as expected. After briefly carrying
out the so called $(3+1)$ decomposition of the Maxwell's equations
generalised to $(d+1)$ dimensional spacetime in section 2 we
investigate the propagation of an electromagnetic wave in vacuum
in section 3. In section 4 the electrostatic oscillations are very
briefly discussed for a cold plasma and the dispersion relations
obtained both for transverse and longitudinal modes. In section 5
we discuss in some detail the propagation of an electromagnetic
wave in cold plasma in the presence of an ambient and homogeneous
magnetic field both parallel and perpendicular to the wave
propagator k. The presence of the magnetic field introduces newer
and interesting modes of oscillation, creating Left and Right
circularly polarized waves, resulting in  the well known classical
phenomenon of Faraday rotation. We also  find that the wave
propagation in the plasma is possible for certain range of
frequencies only and this range critically depends on the number
of dimensions. The paper ends with a short discussion in section
6.
\section*{2.  Field Equations and its Formalism}

  We extend here the (3+1) decomposition of GR as
    formulated by Arnowitt, Deser and Misner (ADM) ~\cite{adm} to a
    higher dimensional space time of (d+1) dimensions. The ADM
    formalism was developed mainly to address the issue associated
    with numerical relativity as also quantization of gravity
    fields, specially when the space time has considerable
    symmetry. This work is connected with plasma physics in curved
    space time. To make use of the intuition from the known results
    of MHD in flat spacetime  it is preferable to spilt the
    ordinary electromagnetic field tensor $F^{\mu\nu}$ into electric and
    magnetic fields $E$ and $B$ in terms of which the field
    equations are more familiar. As the split formalism has been
    extensively discussed and used in the literature ~\cite{Hol,Holc,Zha}
    we shall very briefly touch upon its salient features as
    extended to higher dimensions.

    We define a set of observers ( Fiducial Observers or FIDO ) at
    rest in the space spanned by the hypersurfaces of constant
    universal time, having a d-velocity vector field, $n$
    orthogonal to spatial slices.
    It is well known that for a rotation-free space time ( as we
    are dealing here )

\begin{equation}
n_{i;j}= \sigma_{ij} + \frac{1}{d} ~ \theta \gamma_{ij} - a_{i}
n_{j}
\end{equation}
$(i,j = 1,2,3,\ldots d)$ where $\sigma_{ij}$ is the shear of the
Eulerian world lines given by
\begin{equation}
\sigma_{ij} \equiv \frac{1}{2} \left( n_{i;\mu}\gamma_{j}^{\mu} +
n_{j;\mu}\gamma_{i}^{\mu} \right) - \frac{1}{d} ~\theta\gamma_{ij}
\end{equation}

Here $\gamma_{ij}$ is d-metric spatial tensors
\begin{equation}
a^{i} = n_{;j}^{i} n^{j}
\end{equation}
and
\begin{equation}
\theta = n_{i}^{i} ( = - K )
\end{equation}

are the usual acceleration and expansion scalars and $K$ is the
trace of the extrinsic curvature, $K_{;i}^{i}$.

For our metric we take the ( d+1) dimensional generalized FRW
space time as

\begin{equation}
ds^{2} = dt^{2} - A^{2} \left( dx^{2} + dy^{2} + dz^{2} +
d\psi_{n}^{2}\right)
\end{equation}
(n = 5, 6, 7,  \ldots , d )

where $A \equiv A (t)$ is the scale function. For this particular
metric two relevant quantities $\alpha = \frac{d\tau}{dt}$ ( the
lapse function - the rate of change of fiducial proper time to
that of universal time ) and also the shift vector $\beta$ (
signifying how much spatial co-ordinates are shifted as one moves
from one hypersurface to the other ) naturally reduce to $\alpha =
1$ and $\beta = 0$.

In an earlier work ~\cite{sc1} one of us extensively discussed the
( d+1) dimensional  isotropic and homogeneous space time and
assuming an equation of state, $P = \gamma \rho$ found the scale
factor as ( $P$ = pressure, $\rho$ = energy density )
\begin{equation}
A \sim t^{\frac{2}{d(1+\gamma)}} = t^{n},~~~ n =
\frac{2}{d(1+\gamma)}
\end{equation}
With the extrinsic curvature scalar defined as
\begin{equation}
K = - \theta = - d\frac{\dot{A}}{A} = - d\frac{n}{t}
\end{equation}

we finally write down the Maxwell's equations ~\cite{mac,thor,eva}
(see Mcdonald et al for ( 3+1 ) split for more details )
generalized to ( d+1) dimensions as

\begin{eqnarray}
  \nabla.E &=&4\pi\rho_{e} \\
  \nabla.B &=& 0 \\
  \frac{\partial E}{\partial t} &=& KE + cA^{-1} \nabla \times B - 4\pi J \\
  \frac{\partial B}{\partial t} &=& KB - cA^{-1} \nabla \times E \\
  \frac{\partial \rho_{e}}{\partial t} &=& K \rho_{e} - \nabla .J ~ \textrm{(charge ~ conservation)}
\end{eqnarray}

and finally the particle equation of motion in $(d+1)$ dimensional
as
\begin{equation}
\frac{DA^{d-1}p}{D\tau} =  A^{d-1}q \left(E + A \frac{v}{c}\times
B \right)
\end{equation} or

\begin{equation}
\frac{Dp}{D\tau} = \frac{d-1}{d}Kp + q \left(E + A
\frac{v}{c}\times B \right)
\end{equation}

where
\begin{equation}
\frac{D}{D\tau} = \frac{1}{\alpha} \left(\partial_{t} + v.\nabla
\right)
\end{equation}
is the convective derivative and the d- momentum
\begin{equation}
p = m_{e}\Gamma v
\end{equation}
( $m_{e} $ is the rest mass, $\Gamma$ is the boost factor, and v,
the d-velocity ). We thus see that for the simple metric given by
equation(5)the decomposed ( d+1)- dimensional Maxwell equations
closely mimic the flat space counterparts with some additional
inputs from curved geometry( e.g., A and K terms).

Here $\nabla\cdot$ and $\nabla\times$ are the ordinary Minkowskian
divergence and curl in Cartesian co-ordinates. In what follows we
shall consider, for simplicity, the small amplitude linear theory
such that the convective derivative simply reduces to ordinary
derivative, $\frac{d}{dt}$.

\section*{3. Electromagnetic Waves in Vacuum }

~~~~~~With the set of equations split to ( d + 1) formalism we are
now in a position to attempt applications in varied plasma
phenomena. To start with we discuss briefly the propagation of an
electromagnetic wave in free space in the expanding back ground.
Using equations (9-11) we get via $\nabla\cdot E= 0$ (for vacuum)
the wave equation as
\begin{equation}
    A^{2}\ddot{E} + \left(2d+1 \right)A\dot{A}\dot{E} + \left(d^{2}\dot{A}^{2}+
    dA\ddot{A}\right)E = c^{2}\nabla^{2}E
\end{equation}
Assuming separation of variables in electric field
\begin{equation}
    E (\textbf{x},t ) = E_{t}E_{r}
\end{equation}
we finally get

\begin{equation}
    \ddot{E_{t}}+ (2d+1)\frac{\dot{A}}{A}\dot{E_{t}} +
    \left[d\left(\frac{\ddot{A}}{A}+ d\frac{\dot{A}^{2}}{A^{2}}\right)
    + \frac{k_{i}^{2}c^{2}}{A^{2}}\right]E_{t}= 0
\end{equation}
which, through equation (6) finally reduces to

\begin{equation}
     t^{2}\ddot{E_{t}} + ( 2d + 1) nt\dot{E_{t}} + \left[ dn\left(dn + n -1 \right)
     + k_{i}^{2}c^{2} t^{2(1-n)}\right]E_{t} = 0
\end{equation}

(here $k_{i}$ is a separation constant) corresponding to some
initial fiducial time $t_{i}$. We shall subsequently see that
$k_{i}$ is also identified with the $d$- dimensional wave vector.

On the other hand, as we are dealing with a homogeneous world the
spatial equation remains unchanged yielding a solution
$e^{i(k_{i}\cdot~r)}$ as in special theory of relativity. A little
algebra shows that the time equation is reducible to a Bessel
equation of order $\frac{1}{2}$ as in the 4D case. Thus
dimensionality or the equation of state has apparently no role in
determining the order of the equation. Plugging everything
together we get

\begin{equation}
    \textbf{E} =
    E_{0} \hat{e}t^{\frac{1-(2d+1)n}{2}}H_{\frac{1}{2}}^{(2)}\left[\frac{t^{1-n}}{1-n}k_{i}c
    \right]e^{ik_{i}.r}
\end{equation}

where $H_{\frac{1}{2}}^{(2)}$ is a Hankel function of order
$\frac{1}{2}$. Replacing the asymptotic form of $H_{\frac{1}{2}}$
we get

\begin{equation}
    E =
    E_{0}i \sqrt{\frac{2k_{i}cd(1+\gamma)}{\pi\{d(1+\gamma)-2\}}}t^{-\frac{2}{1+\gamma}}e^{-\frac{ik_{i}cd(1+\gamma)}{d(1+\gamma)-2}
    t^{\frac{d(1+\gamma)-2}{d(1+\gamma)}}}e^{ik_{i}.r}
 \end{equation}

It represents a d-dimensional `damped' harmonic wave as one
encounters in mechanical vibration. While in mechanical motion the
damping occurs due to friction here the expansion of the universe
seemingly causes some sort of damping. For pre recombination era
in $(d+1)$ dimensional space time $\gamma = \frac{1}{d} $ and the
damping factor is  $A^{-d}$. Hence the damping of the wave
amplitude apparently decreases with number of dimensions. This
finding merits some explanation. We have remarked earlier that the
amplitude decay is somewhat geometrical in nature caused by the
expansion and curvature. In that case as with dimension the
expansion rate decreases one expects that damping should be larger
in 4D but a little inspection of the last relation shows that when
we plug in the expression of `$A$' (equation (6)) the last
relation further reduces to $t^{-\frac{2d}{d+1}}$. So the damping
actually increases in higher dimensional spacetime. On the other
hand, for a fixed $d$ the damping factor is $t^{-2}$ for post
recombination era ($\gamma=0$) (see figure 1). Alternatively for
the case of a very large number of dimensions the damping
asymptotically reaches $t^{-2}$, a form set for post recombination
era. It has not escaped our notice that the scaling of $E$ or $B$
for $\gamma=0$ is independent of the number of dimensions unlike
the radiation case. So the amplitude factor gets increasingly
damped as the universe ages. The fact that damping of the $E$ or
$B$ increases with the number of dimensions in the early universe
has a number of interesting theoretical implications. Firstly we
mention in the introduction that in order that the universe
evolves isotropically according to the FRW model there should be
efficient mechanisms to damp the primordial magnetic field as
early as possible. In that respect the HD spacetime has some
inherent advantage over the standard 4D in the sense that the
damping is faster in HD. Secondly the magnetic field at small
scales may influence the bigbang nucleosynthesis and change the
primordial abundances of light elements by significantly changing
the expansion rate of the universe at the corresponding time. The
success of the standard BBN scenario can provide an interesting
set of bounds on the intensity of the magnetic field at that epoch
~\cite{giovanni}, indirectly constraining the number of dimensions
of the spacetime. At this stage it may not be out of place to call
attention to the fact that most of the above findings are of
theoretical nature only and it is not feasible  to relate them to
current astrophysical data. Because multidimensional cosmological
models lose much of their relevance well before the onset of
bigbang nucleosynthesis and current observational findings can be
explained for all practical purposes if the cosmological evolution
be modelled along the standard four dimensional spacetime.

If as usual we set $k_{i}c = \omega_{i}$ ( the angular frequency
of the wave at some initial time $t = t_{i}$ ) then the above
equation may be rewritten as

\begin{figure}[h]
\begin{center}
  \includegraphics[width=6cm]{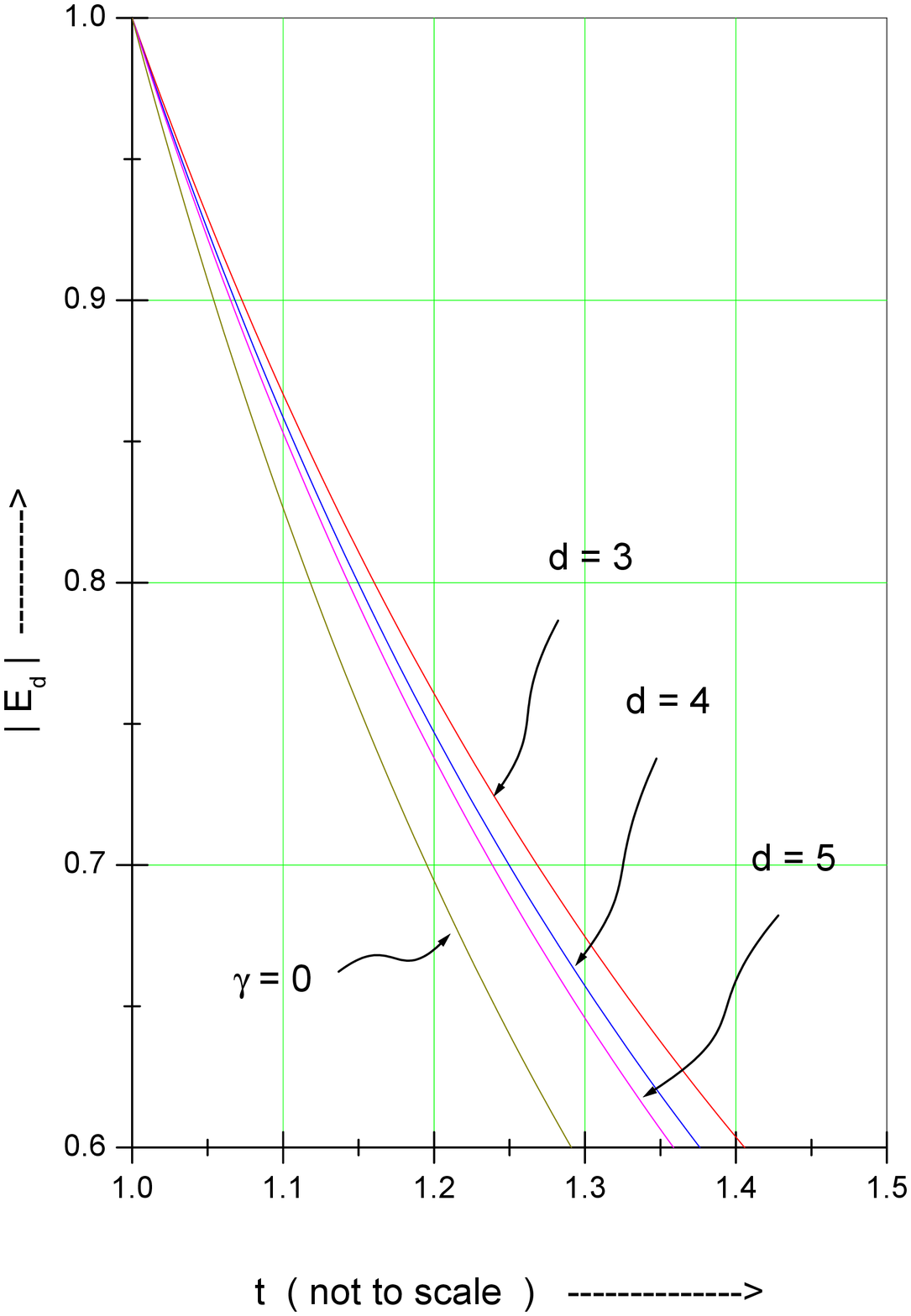}
  \caption{
 \small \emph{ $\mid E_{d} \mid \sim t  $ graph}
}
\end{center}
\end{figure}
\begin{eqnarray}
    E &=& E_{0}i \sqrt{\frac{2k_{i}cd(1+\gamma)}{\pi\{d(1+\gamma)-
    2\}}}t^{-\frac{2}{1+\gamma}}e^{-i\omega_{i}t\frac{d(1+\gamma)}{d(1+\gamma)-2}
    t^{-\frac{2}{d(1+\gamma)}}}e^{ik_{i}.r}\nonumber \\
    &=& E_{0}i \sqrt{\frac{2k_{i}cd(1+\gamma)}{\pi\{d(1+\gamma)-2\}}}t^{-\frac{2}{1+\gamma}}e^{- i \frac{d(1+
    \gamma)}{d(1+\gamma)-2}\omega_{d}t}e^{ik_{i}.r}
\end{eqnarray}
where
\begin{equation}
    \omega_{d} = \omega_{i}t^{-\frac{2}{d(1+\gamma)}}
\end{equation}
gives a measure of the red shift of the photon due to background
expansion. For radiation dominated era $\gamma = \frac{1}{d}$ ,
$\omega_{d} = \omega_{i}t^{-\frac{2}{d+1}}$, so the rate at which
the frequency decreases is maximum in 4D universe. Moreover
damping is greater in radiation era (see figure 2).
\begin{figure}[h]
\begin{center}
  \includegraphics[width=6cm]{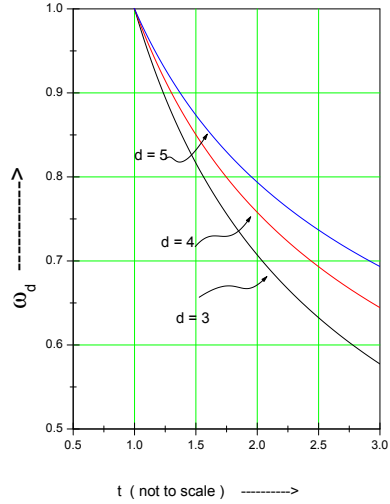}
  \caption{\small \emph{ $\omega _{d}  \sim t$ graph. As the
number of dimension increases red shift decreases}
    }
\end{center}
\end{figure}

Returning again to the equation (22) we see that the horizon for
our metric is given by

\begin{equation}
    L_{d} = \int\frac{cdt}{t^{\frac{2}{d(1+\gamma)}}}
     = \frac{d(1+\gamma)}{d(1+\gamma)-2} c t^{\frac{d(1+\gamma)-2}{d(1+\gamma)}}
\end{equation}

so the equation may be recast as

\begin{equation}
    E = E_{0}(x,t)e^{i({k.r-k_{i}.L_
    {d}})}=E_{0}(x,t)e^{ik_{i}(\hat{k}x-L)}
\end{equation}

exactly similar to the Newtonian result where the horizon is
simply $L=ct$.
\section*{4. Electrostatic Oscillation}

In this section we shall very briefly consider an electromagnetic
wave in a 2-component plasma. For simplicity we assume a small
amplitude wave such that, $ v \times B=0$ because it is of second
order in perturbed quantities.  This, in turn, allows us to
neglect the motion of ions. Skipping intermediate mathematical
steps for space we get via equations (14) and (16) for our metric
(5)
\begin{equation}
v=\frac{iqt^{\frac{2}{d(1+\gamma)}}}{m_{e} \Gamma \omega_{i}}E = -
\frac{ie}{m_{e} \Gamma \omega_{d}}E
\end{equation}
The last equation is very similar to the flat space case except
that here, $\omega_{d}$ is not a constant but shares the
background expansion. When one uses the last equation in Maxwell
equation (10) through $ J= n_{0}q v$ we arrive at Coulomb's law
\begin{equation}
\nabla\times B = -\frac{i\omega_{i}}{c}\epsilon E
\end{equation} where
\begin{equation}
\epsilon (\omega_{d}) = 1 - \frac{\omega_{pT}^{2}}{\omega_{d}^{2}}
\end{equation}
is the dielectric constant of the plasma medium with the suffix
$'T'$ signifying transverse mode. The $\omega_{pT}^{2}$ is related
to the well known plasma frequency ~\cite{yu},
\begin{eqnarray}
\omega_{p}^{2}&=& \frac{b_{d}
n_{0}q^{2}}{m_{e}}~,~~~\omega_{pT}^{2}\sim\frac{\omega_{p}^{2}}{\Gamma} \nonumber \\
 b_{d} &= & \frac{2^{\frac{d}{2}}\pi^{\frac{d}{2}}}{(d-1)!!} ~~~~~~~~~~~~ \mathrm{(d~~~ even)}
 ,\nonumber \\
 &=& \frac{2^{(\frac{d+1)}{2}}\pi^{\frac{(d-1)}{2}}}{(d-2)!!} ~~~~~~~~~\mathrm{(d~~ odd)}\nonumber
\end{eqnarray}
 Skipping mathematical details it
can also be shown that the transverse and the longitudinal modes
give the dispersion relations as
\begin{equation}
\omega_{T}^{2} = \omega_{p}^{2} + c^{2}k^{2}
\end{equation}

\begin{equation}
\omega_{L}^{2} = \omega_{p}^{2}
\end{equation}
These relations are pretty well known in the special relativistic
case, excepting that here all the quantities depend on time as
well as the total number of dimensions.

Apparently the equation (29) has the same Newtonian form but both
the frequencies depend on the scale factor, `$A$' which, again, is
a function of both the number of dimensions and the equation of
state chosen.To end the section let us investigate the time
dependence of the dielectric constant in equation(29). Now,
$\omega_{pT}^{2}$ should share the time evolution of the
background electron number density, $n_{0}$ ( the inverse of
volume of the universe) i.e., $ n_{0} \sim A^{-d}\sim
t^{-\frac{2}{1+\gamma}}$. Again, from (27) we get, $v\Gamma\sim
A^{n(1-d)}$. The equation (10) further dictates that $v\sim
A^{-1}$, which gives $\Gamma\sim A^{(2-d)}$. So, $\omega_{pT}^{2}
\sim \frac{n_{0}}{\Gamma}\sim A^{-2}\sim
t^{-\frac{4}{d(1+\gamma)}}$. On the other hand the equation (24)
implies that $\omega_{d}^{2} \sim A^{-2}$, hence
$\epsilon$($\omega_{d}$) does not explicitly depend on time. This
is a remarkable result in the sense that for a FRW type of metric
the dielectric constant is a real constant irrespective of not
only the total number of dimensions but also on the equation of
state i.e., $\epsilon$($\omega_{d}$) continues to remain constant
all through the evolution of the universe.


\section*{5. Electromagnetic Oscillations in Cold Plasma}

In this section we investigate the situation where a plasma in
thermodynamic equilibrium is slightly disturbed through the
passage of an electromagnetic wave. We assume that an external
ambient magnetic field is also present. We, however, assume the
plasma medium to be cold so that the pressure can be neglected
when considering the particle equation of motion. In stellar
systems one often encounters situations where relaxation times are
much larger than the age of the universe so that collisions (
hence pressure ) may be neglected. The effect of an electric field
is not generally seriously considered because of the well known
Debye shielding effect. The general problem of an electromagnetic
wave propagating along an arbitrary direction with the external
magnetic field is given by Appleton and Hartee in the Newtonian
case when studying the propagation of radio waves in ionosphere.
Holcomb ~\cite{Holc} studied in FRW metric a specialised situation
of the A-H equation in the dust case. Considering the fact that a
general solution with arbitrary $\theta$ is very difficult to
tackle in an expanding background with arbitrary number of
dimension we shall restrict ourselves to the cases when the
electromagnetic wave propagates parallel and perpendicular to the
magnetic field. However the topic is of great importance in
astrophysics and space science where electromagnetic wave
propagation in magnetized plasma is very relevant.

\vspace{.3 cm}

\textbf{ Case I ($\vec{B}\parallel \textbf{k}$) :}

we assume that the external, uniform magnetic field and the wave
vector
           $\mathbf{k}$ are both aligned along the $i^{th}$ direction
            (say z direction with $i = 3$) in the $d$- dimensional space, such that
           $\textbf{k} = |k |\mathbf{e_{z}}$ and $\mathbf{B} = | B|\mathbf{e_{z}}$.
           As is customary in the analogous 3-dimensional static
           space we also assume  that all the perturbed quantities have
           the same time dependence given by equation (23) such that
           the linearized equation of motion (13) takes the form

\begin{equation}
   i\omega_{d}m_{e} v = e \left(E + A\frac{v}{c}\times B
\right)
\end{equation}
[ Here E is $\perp$r to $k$ and considering that we are dealing
with a $(d+1)$ dimensional space time it has components $E_{1}$,
$E_{2}$, $E_{3}$, $E_{4}$, \ldots , $E_{j}$ .
]\\
Replacing $\frac{\partial E}{\partial t}$ via equation (23) by

\begin{equation}
   \frac{\partial E}{\partial t} = - \left[i\omega_{d} +
    \frac{2}{(1+\gamma)t}\right]E
\end{equation}

which, when plugged in equation (10) gives, after a long but
fairly straight forward calculation gives  for j=1

\begin{eqnarray*}
   \left( \nabla\times B \right)_{1} &=& - \frac{i\omega_{d}}{c}A
   \left[ \left(1 - \frac{\omega_{p}^{2}}{\omega_{d}^{2}} \right)
   E_{1} - \frac{\omega_{p}^{2}}{\omega_{d}^{2} - \omega_{c}^{2}}
   \frac{\omega_{c}^{2}}{\omega_{d}^{2}}E_{1}
   +i \sum_{j=2}^{d}\frac{\omega_{p}^{2}}{\omega_{d}^{2} - \omega_{c}^{2}}\frac{\omega_{c}}{\omega_{d}}E_{j}
   \right]
   \end{eqnarray*}
\begin{equation}
   =
   - i\frac{\omega_{d}}{c} \left[\left(1 -
  \frac{\omega_{p}^{2}}{\omega_{d}^{2}-\omega_{c}^{2}}
   \right)E_{1} +
    i \sum_{j=2,j\neq 1}^{d}\frac{\omega_{c}}{\omega_{d}}\frac{\omega_{p}^{2}}{\omega_{d}^{2} - \omega_{c}^{2}}E_{j}
   \right]
\end{equation}

For the case $j = 3$ (i.e., along the direction of the magnetic
field ) it takes a simple form
\begin{equation}
  \left( \nabla \times B \right)_{3} = - i
  \frac{\omega_{d}}{c}
  \left(1-\frac{\omega_{p}^{2}}{\omega_{d}^{2}}\right)E_{3}
\end{equation}
  repeating the process for the remaining $(d-2)$ components
  we can write for the $\mu ^{th}$ component a tensorial relation as
\begin{equation}
  \left( \nabla \times B \right)_{\mu} = - i
  \frac{\omega_{d}}{c}\epsilon_{\mu\nu}E_{\nu}
\end{equation}
($\mu,\nu = 1, 2, 3, \ldots , d $). where $\epsilon_{\mu\nu}$ is
rank 2 skew symmetric tensor of order `d'. A little inspection
shows that
\begin{eqnarray}
  \epsilon_{11} &=& \epsilon_{22}=\epsilon_{44}=\epsilon_{55} =  \ldots
  = \epsilon_{dd}=1 - \frac{\omega_{p}^{2}}{\omega_{d}^{2}-\omega_{c}^{2}}= p_{1} \textrm{(say)}\\
  \epsilon_{12} &=& \epsilon_{14}=\epsilon_{15}= \ldots = \epsilon_{1d}
   =\frac{\omega_{c}}{\omega_{d}} \frac{\omega_{p}^{2}}{\omega_{d}^{2}-\omega_{c}^{2}} = p_{2}\\
  \epsilon_{31} &=& \epsilon_{32}= \epsilon_{34}= \ldots = \epsilon_{3d} = 0 \\
  \epsilon_{33} &=& 1 - \frac{\omega_{p}^{2}}{\omega_{d}^{2}} = p_{3}
\end{eqnarray}
so the $(d\times d)$ permitivity tensor comes out to be

\begin{equation}
\epsilon_{\mu\nu} =\left(%
\begin{array}{cccccc}
  p_{1} & ip_{2} & 0& ip_{2} & . & ip_{d} \\
  -ip_{2} & p_{1} & 0 & ip_{2} & . & 0 \\
  0 & 0 & p_{3} & 0& . & 0 \\
  -ip_{2} & -ip_{2} & 0 & p_{1} &. & 0 \\
  . & . & . & . & . & . \\
  -ip_{d} & . & . &. & . & p_{1} \\
\end{array}%
\right)
\end{equation}

Here $\omega_{p}$ is the plasma frequency given by

\begin{equation}
\omega_{p}^{2} = \frac{b_{d} n_{0}e^{2}}{m_{e}}
  \end{equation}
and the electron cyclotron frequency is given by
 \vspace{2 cm}
\begin{figure}[h]
\begin{center}
  \includegraphics[width=10 cm]{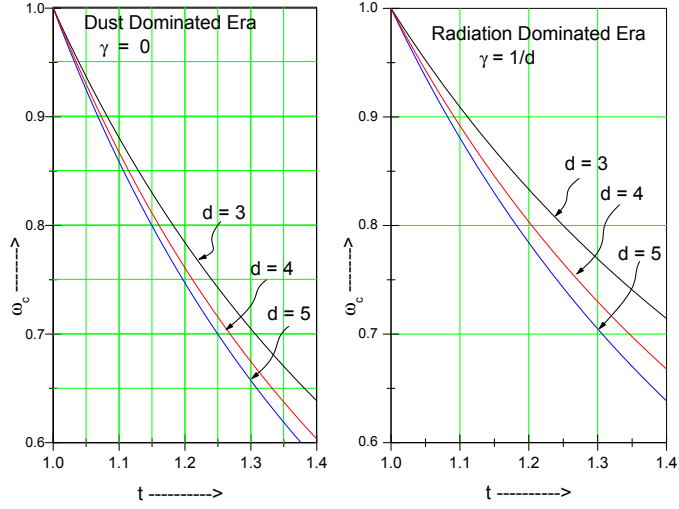}
 \vspace{-2.5 cm}
  \caption{
\small \emph{ $\omega _{c} \sim t$ graph for both radiation and
matter dominated era. As the number of dimension increases $\omega
_{c}$ decreases. Further the decay is sharper in dust case
compared to radiation case.
  }
    }
\end{center}
\end{figure}
\begin{equation}
\omega_{c} = \frac{eB}{m_{e} c}t^{\frac{2}{d(1+\gamma)}}\equiv
\frac{e\hat{B}}{m_{e} c}
\end{equation}
where $\hat{B}$, the orthogonal magnitude of the ambient magnetic
field is given by $\hat{B}=|(B)_{z}(B)^{z}|^{1/2} = B
t^{\frac{2}{d(1+\gamma)}}$ for our system (see figure 3).

If the magnetic field is switched off $(\omega_{c}=0)$ the
equation (34) reduces to

\begin{equation}
  \left(\nabla \times B\right)_{j} = - \frac{e\omega_{i}}{c}\epsilon E_{j}
\end{equation}
exactly similar to the expression(28) of the section 4. Thus the
introduction of the magnetic field generates varied modes
transforming the dielectric constant scalar $\epsilon$ in equation
(35) to a second rank tensor $\epsilon_{ij}$. Although the
equations (30) - (40) exactly resemble the analogous expressions
in Newtonian theory the fact remains that that all the frequencies
now depend on time rather than being constant. Further the
cyclotron frequency $\omega_{c}$ decays as
$t^{-\frac{2(d-1)}{d(1+\gamma)}}$ exactly similar to the
orthogonal component of the magnetic field.

If we take the curl of the equation (36) and replace $\nabla$ by
$ik\hat{e_{z}}$, after a long but fairly straight forward
calculation we are led to the matrix form

\begin{equation}
\left(%
\begin{array}{cccccc}
  (1-\frac{p_{1}}{\mathbf{n}^{2}}) & -i\frac{p_{2}}{\mathbf{n}^{2}} & 0 & 0 & . & 0 \\
  i\frac{p_{2}}{\mathbf{n}^{2}}  & (1-\frac{p_{1}}{\mathbf{n}^{2}}) & 0 & 0 &  . & 0 \\
  0 & 0& -\frac{p_{3}}{\mathbf{n}^{2}} & 0 & . & 0 \\
  . & . & . & . & . & . \\
  .& . & . & . & . & . \\
  0 & 0 & 0 &0 & . & (1-\frac{p_{1}}{\mathbf{n}^{2}}) \\
\end{array}%
\right)
\left(%
\begin{array}{c}
  E_{1} \\
  E_{2} \\
  E_{3}\\
  . \\
  . \\
   E_{d} \\
\end{array}%
\right) = \left(%
\begin{array}{c}
  0 \\
  0 \\
  0 \\
  . \\
  .\\
  0 \\
\end{array}%
\right)
\end{equation}

where \begin{equation}\mathbf{n}^{2}=
\frac{c^{2}k^{2}}{\omega_{i}^2}
\end{equation}
 where $\mathbf{n}$
is the refractive index of the plasma medium.

 Three modes are
possible.

First the longitudinal mode characterized by $E_{1}= E_{2}= E_{4}=
\ldots =E_{d} = 0$, ($ E_{3} \neq 0$ and $p_{3}=0$). Since the
displacement is along $Z$ direction the magnetic field has no role
to play and $\omega_{d} = \omega_{p}$.

As we are more interested in the dynamics of the electromagnetic
waves rather than the plasma oscillation as such we take $E_{3}
=0$ in equation (45). Setting the determinant of the resulting
$(d-1)\times (d-1)$ matrix in  equation (45) to zero we get

\begin{equation}
  \mathbf{n}^{2} = p_{1}\pm p_{2}
\end{equation}

The plus sign gives via equation (36)

\begin{equation}
  E_{\mu}- \emph{i}E_{\nu} = 0  ~~~~~(\mathrm{\mu \neq \nu \neq 3}
  ~~~,~~~ \emph{i} = \sqrt{-1} )
\end{equation}

such that
\begin{equation}
E_{l} = \left(\hat{e_{\mu}} -\emph{i}\hat{e_{\nu}}\right)
e^{i(k_{L}z-\omega_{d}t)}
\end{equation}

corresponding to left circularly polarized wave.

The wave number $k_{L}$ can be found from equations (37, 38, 46,
47) as
\begin{equation}
k_{L} =
\frac{\omega_{d}}{c}\left[1-\frac{\omega_{p}^{2}}{\omega_{d}(\omega_{d}
+ \omega_{c}) }\right]^{1/2}
\end{equation}

On the other hand for the minus sign in  (47) we get

\begin{equation}
E_{R} = \left(\hat{e_{\mu}} + \hat{e_{\nu}}
\right)e^{[\emph{i}(k_{R}z-\omega t)]^{1/2}}
\end{equation}
representing RCP wave with

\begin{equation}
k_{R} = \frac{\omega_{d}}{c} \left[1-
\frac{\omega_{p}^{2}}{\omega_{d}(\omega_{d} -
\omega_{c})}\right]^{1/2}
\end{equation}

It is clear that the two eigen modes have different phase and
group velocities and unlike the former case the Right Circularly
Polarized (RCP) wave has a resonance at $\omega = \omega_{f} =
\omega_{c}$ where the phase velocities vanish. The expressions so
far exactly resemble the ones found in the propagation of an
electromagnetic wave with an ambient magnetic field in Newtonian
mechanics. However, here all the quantities $\omega_{d}$,
$\omega_{p}$ etc. depend on time, a dependence modelled by the
form of line-element, the number of dimensions and also the
equation of state.

  It should be noted that the time dependence of $\omega_{d}$ and
  $\omega_{c}$ are different being $\omega_{d} \sim
  t^{-\frac{2}{d(1+\gamma)}}$ and  $\omega_{c} \sim
  t^{-\frac{2(d-1)}{d(1+\gamma)}}$. So there is no fixed resonant
  frequency as in Newtonian case but with time it changes.

It also follows from equation (52) that for
\begin{equation}
\omega_{d} = \omega_{1} = \frac{1}{2} \left[ \omega_{c} +
\sqrt{\omega_{c}^{2} + 4 \omega_{p}^{2}}  \right]
\end{equation}
the wave vector $k$ vanishes and our analysis breaks down. So the
wave propagates for $\omega_{d} < \omega_{c}$ and $\omega_{1} <
\omega_{d}$, otherwise it becomes evanescent. Moreover, as the
temporal dependence of $\omega_{c}$ and $\omega_{d}$ are different
the magnitude of the allowed region changes. With dimensions
$\omega_{c}$ decays more sharply than $\omega_{d}$. Thus the
propagation of the electromagnetic wave is more restricted in
higher dimensions than the usual 4D.

Returning to the Left Circularly Polarized (LCP) wave we see that
the wave vector vanishes for

\begin{equation}
\omega_{d} = \omega_{2} = \frac{1}{2} \left[ -\omega_{c} +
\sqrt{\omega_{c}^{2} + 4 \omega_{p}^{2}}  \right]
\end{equation}

and so the wave propagates for $\omega_{d} > \omega_{2}$

From what has been discussed above it is tempting to look for
Faraday rotation (see ref. 4 for recent astrophysical data)
analogous to the Newtonian case. Assuming that an electromagnetic
wave traverses a distance $z$ in a plasma medium with a magnetic
field subject to the restriction on frequencies discussed above
the Faraday rotation is given by

\begin{equation}
\theta = \frac{k_{L} - k_{R} }{2}Z
\end{equation}

It should be noted that one should revert to the physical
co-ordinate rather than the co-moving one we are considering here.
Accordingly $Z_{ph}= t^{-\frac{2}{d(1+\gamma)}}$, $Z_{cm}$ and
$\theta$ finally comes out to be via equation (23)

\begin{equation}
\theta = \frac{k_{Li} - k_{Ri} }{2}Z_{ph}
\end{equation}
with no dependence on time. so apparently the number of dimensions
and the equation of state have no impact on this classical result.
It may be relevant to mention that measurements of the radio waves
from the extra galactic sources  suggest that various spiral
galaxies are endowed with magnetic fields whose intensities are of
the same order of magnitude as that of Milky way ~\cite{Kronberg}
i.e., of the order of microgauss corresponding to an energy
density stored today in CMBR energy spectrum peaked around a
frequency of 30 GHz.

\vspace{.5 cm}

\textbf{ Case II ($\vec{B}\bot \vec{k}$) :}

Let us very briefly consider the case of a plasma with a uniform
magnetic field $\textbf{B} = B_{0}\hat{e}_{z}$, through which an
electromagnetic wave is propagating with propagation vector
$\vec{\textbf{k}}= k\hat{e_{x}}$, perpendicular to the magnetic
field. Here two modes are possible. As the mathematical exercise
closely resembles the case I we shall totally skip intermediate
steps  to write the final form as
 \vspace{.2 cm}

  1. First mode (called
ordinary wave)  with displacements in z direction ( i.e.
$\parallel B$) having the dispersion relation

\begin{equation}
\omega_{d}^{2} = \omega_{p}^{2} + k^{2}c^{2}
\end{equation}
 as the magnetic field has no influence for motion parallel to
 itself the equation (57) is exactly same as equation (30) for the
 electrostatic oscillation.
\vspace{.2 cm}

2. Second mode (called extraordinary wave) with displacements in
(d-1)-dimensional hypersurface  ($\bot B $) having dispersion
relation

\begin{equation}
\frac{c^{2}k^{2}}{\omega_{d}^{2}} = 1 -
\frac{\omega_{p}^{2}}{\omega_{d}^{2}} \frac{\omega_{d}^{2} -
\omega_{p}^{2}}{\omega_{d}^{2} -\omega_{p}^{2} - \omega_{c}^{2}}
\end{equation}

Before ending the section a final remark may be in order. We know
that pulsars are rotating neutron stars giving out pulses of radio
waves periodically, which are affected by the interstellar medium
during their propagation to reach us. If the interstellar medium
has a component of magnetic field parallel to propagation
direction then as shown earlier the plane of polarization will
suffer Faraday rotation depending on frequency, having a spread in
the rotation angle. This spread may have, \emph{in principle at
least}, some imprint on the nature of expanding universe.
\section*{6. Discussion}
With the help of (3+1) formalism the Einstein-Maxwell and the
electrodynamical equations are written for a (d+1) dimensional
FRW-like spacetime in presence of plasma and linearised equations
are solved for different phases of the universe.The analysis
essentially generalises to HD the well known results of Holcomb
and Tajima. The salient features of our analysis may be summarised
as:
 \vspace{.1 cm}

1. For a propagating wave in HD in vacuum the photons redshift
most in 4D and for a fixed $d$ in radiation dominated model.
\vspace{.1 cm}

2. Although the plasma is sharing the expansion of the background
the dielectric constant remains a true constant. So the photons
are in a sense self similar. This result was found earlier by
Holcomb and Tajima. We here generalise this remarkable result to
the case of extra dimensional spacetime and also for a fluid
obeying  a general equation of state. It may be tempting to
suggest that the fact that the classical flat space result of the
constancy of the dielectric constant is carried over to non static
curved background and that too in higher dimensions may be due to
the conformal flatness of the particular metric analysed here. So
one should move with caution against any far fetched
generalisation and in other complicated space time this result may
not be true.
 \vspace{.1 cm}

3. In the presence of an external magnetic field many interesting
oscillation modes manifest themselves. A simplified
Appleton-Hartee type of solution generalised to higher dimensions
is obtained in curved spacetime. Only a selected range of
frequencies are available for propagation here.
\vspace{.1 cm}

4. The well known phenomenon of Faraday rotation is obtained.
\vspace{.1 cm}

To end  a final remark may be in order. The present work suffers
from two serious disqualifications. For sake of mathematical
simplicity we work out everything assuming a linearised plasma
theory. Conditions under which one may assume linearized plasma
theory may well exist in Newtonnian theory, but we are not being
able to clearly formulate those things for the case of a nonflat
spacetime and that too when it is expanding. Secondly most
observational evidences suggest that even if one starts with a
higher dimensional phase the universe underwent the self
compactification transition much earlier than the epoch when the
big bang nucleosynthesis sets in. So although literature abounds
with works (for example, higher dimensional black holes and its
thermodynamics etc.) studying the standard electromagnetic as well
as MHD laws in the framework of multidimensions becomes a sort of
\emph{suspect }. In that sense our analysis is more of a purely
theoretical nature without much direct physical applications. In
future work one should try to generalise these results in the
realm of non linear plasma and also attempt to relate some of our
findings to known astrophysical data.

\textbf{Acknowledgment : } One of us(SC) acknowledges the
financial support of UGC, New Delhi for carrying out this work.


\begin{thebibliography}{40}
\bibitem{prl}Pisin Chen and Kwang- Chang Lai, 2007 \emph{Phys. Rev. Lett.} \textbf{99}
 231302
 \bibitem{smoot}G F Smoot, 2007 \emph{Rev. Mod. Phys.} \textbf{79} 1370
\bibitem{bb}S Banerji and A Banerjee, 2007 \emph{General Relativity and Cosmology},
 Elsevier
\bibitem{Kronberg}P P Kronberg,  1994 \emph{Rep.Prog.Phy} \textbf{57} 325
\bibitem{giovanni}M Gasperini, M Giovanini and G Veneziano,1995  \emph{Phy. Rev.Lett.}
\textbf{75} 3796
\bibitem{beck}R Beck, A Brandenberg, D Moss, A A Shukurov and D.
Sokoloff,1996  \emph{Annu. Rev. Astron. Astrophys.}\textbf{34} 155
\bibitem{barrow}C G Tsagas and J D Barrow ,  1997 \emph{Class. Quant. Grav.}
\textbf{14} 2539
\bibitem{zeldovich}Y B Zeldovich, A A Ruzmaikin and  D D Sokoloff,
1983  \emph{Magnetic Fields in Astrophysics}, ( Mcgraw Hill), N Y

\bibitem{pr}D Hooper and S Profumo, `Dark Matter and Collider
 Phenomenology of Universal Extra Dimensions',  2007  \emph{Physics Reports }
  \textbf{453} p 27 - 115
  \bibitem{gr}S Chatterjee, A Banerjee and Z H Zhang, 2006 \emph{Int. J. Mod. Phys.}
\textbf{A21} 4035 ; N Banerjee and S Das,  2006 \emph{Mod. Phys.
Lett.} \textbf{A 21} 2663
  \bibitem{subenoy} U Debnath, A  Banerjee and S  Chakraborty ,  2004 \emph{Class. Quant. Grav.}
\textbf{21} 5609
\bibitem{wesson}P S Wesson,  1999 \emph{Space Time Matter }( World Scientific),
Singapore
\bibitem{Hol}K  A  Holcomb and T Tajima,  1989 \emph{Phy. Rev.} \textbf{D40} 3809
\bibitem{Holc}K A Holcomb,  1990 \emph{Astrophysical. J.} \textbf{362} 381
\bibitem{absc} A  Banerjee, S  Chatterjee, A  Sil and N  Banerjee,  1994 \emph{Phy. Rev.}
\textbf{D50} 1161
 \bibitem{dettmann}C  P  Dettmann, N  E Frankel, 1993 \emph{Phys. Rev.} \textbf{D48} 5655
 \bibitem{sc1}S Chatterjee and B Bhui,  1990 \emph{Mon. Not. R. Astron. Soc.} \textbf{247} 57
 \bibitem{adm}R Arnowitt, S Deser and C W Misner, in Gravitation: 1962  \emph{An Introduction to current Research},
edited by L witten ( Wiley)  New York
\bibitem{Zha}X H Zhang, 1989 \emph{Phy. Rev.} \textbf{D39} 2933
\bibitem{mac}D A Macdonald and K Thorne,  1982 \emph{Mon. Not. R. Astron. Soc.} \textbf{198} 345
\bibitem{thor}K Thorne and D A Macdonald, 1982 \emph{Mon. Not. R. Astron. Soc.} \textbf{198} 339
\bibitem{eva}C R Evans and J F Hawley,  1988 \emph{Astrophysical J.} \textbf{332} 659
\bibitem{yu}V  M Emelyanov, Yu P  Nikitin, I L Rozenbal and A V Berkov
, 1996 \emph{Phys.Rep.} \textbf{143} p 1-68
\end{thebibliography}
\end{document}